\documentclass[3p,onecolumn]{elsarticle}

\usepackage{hyperref}
\usepackage{txfonts}
\usepackage{natbib}
\usepackage{graphicx}
\usepackage[utf8]{inputenc}
\usepackage{multirow}

\journal{Icarus}

\biboptions{authoryear}

\def\geomalb{p$_{\mathrm{V}}$}

\def\ggg{G!k\'un$||$'h\`omd\'im\`a}

\newcommand{\ky}{2002\,KY$_{14}$}
\newcommand{\kylong}{(250112)\,2002\,KY$_{14}$}
\newcommand{\plshort}{2010\,PL$_{66}$}
\newcommand{\pllong}{(499522)\,2010\,PL$_{66}$}
\newcommand{\ph}{2013\,PH$_{44}$}
\newcommand{\phlong}{(471931)\,2013\,PH$_{44}$}
\newcommand{\vu}{2012\,VU$_{85}$}
\newcommand{\vulong}{(463368)\,2012\,VU$_{85}$}
\newcommand{\jj}{2010\,JJ$_{124}$}
\newcommand{\jjlong}{2010\,JJ$_{124}$}
\newcommand{\gx}{2010\,GX$_{34}$}
\newcommand{\gxlong}{2010\,GX$_{34}$}
\newcommand{\yd}{2009\,YD$_7$}
\newcommand{\ydlong}{(353222)\,2009\,YD$_{7}$}
\newcommand{\aeshort}{2016\,AE$_{193}$}
\newcommand{\aelong}{(514312)\,2016\,AE$_{193}$}
\newcommand{\cxshort}{2017\,CX$_{33}$}
\newcommand{\cxlong}{(523798)\,2017\,CX$_{33}$}
\newcommand{\fzshort}{2015\,FZ$_{117}$}
\newcommand{\fz}{2015\,FZ$_{117}$}
\newcommand{\fzlong}{(472760)\,2015\,FZ$_{117}$}

\newcommand{\mruncorr}{m$_{11}^\mathrm{R}$}
\newcommand{\mrcorr}{m$_{110}^\mathrm{R}$}

\begin{document}

\newcommand{\arcmin}{$^{\prime}$}
\newcommand{\arcsec}{$^{\prime\prime}$}
\newcommand{\degr}{$^{\circ}$}
\newcommand{\mjysr}{MJy\,sr$^{-1}$}
\newcommand{\dr}{2012\,DR$_{30}$}
\newcommand{\angstrom}{\mbox{\normalfont\AA}}
\newcommand\sun{\hbox{$\odot$}}
\def\lesssim{\mathrel{\hbox{\rlap{\hbox{\lower3pt\hbox{$\sim$}}}\hbox{\raise2pt\hbox{$<$}}}}}
\def\degr{\hbox{$^\circ$}}
\def\arcmin{\hbox{$^\prime$}}
\def\arcsec{\hbox{$^{\prime\prime}$}}
\def\utw{\smash{\rlap{\lower5pt\hbox{$\sim$}}}}
\def\udtw{\smash{\rlap{\lower6pt\hbox{$\approx$}}}}
\def\fd{\hbox{${.}\!\!^{\rm d}$}}
\def\fh{\hbox{${.}\!\!^{\rm h}$}}
\def\fm{\hbox{${.}\!\!^{\rm m}$}}
\def\fs{\hbox{${.}\!\!^{\rm s}$}}
\def\fdg{\hbox{${.}\!\!^\circ$}}
\def\farcm{\hbox{${.}\mkern-4mu^\prime$}}
\def\farcs{\hbox{${.}\!\!^{\prime\prime}$}}
\def\fp{\hbox{${.}\!\!^{\scriptscriptstyle\rm p}$}}

\sloppy

\begin{frontmatter}
\title{Light curves of ten Centaurs from K2 measurements}
\author[csfk,elteiop]{G\'abor~Marton} 
\author[csfk,elteiop]{Csaba~Kiss} 
\author[csfk,lendulet]{L\'aszl\'o Moln\'ar}
\author[csfk]{Andr\'as P\'al}
\author[csfk,elte]{Anik\'o Farkas-Tak\'acs}
\author[got,exoplanet]{Gyula M. Szab\'o}
\author[mpe]{Thomas M\"uller}
\author[mpe]{Victor Ali-Lagoa}
\author[csfk,lendulet]{R\'obert Szab\'o}
\author[csfk]{J\'ozsef Vink\'o}
\author[csfk]{Kriszti\'an S\'arneczky}
\author[csfk,elte]{Csilla E. Kalup}
\author[amu]{Anna Marciniak}
\author[gran]{Rene Duffard}
\author[csfk,syd]{L\'aszl\'o L. Kiss}

\address[csfk]{Konkoly Observatory, Research Centre for Astronomy and Earth Sciences, Konkoly Thege 15-17, H-1121~Budapest, Hungary}
\address[elteiop]{ELTE E\"otv\"os Lor\'and University, Institute of Physics, P\'azm\'any P. st. 1/A, 1171 Budapest, Hungary}
\address[lendulet]{MTA CSFK Lend\"ulet Near-Field Cosmology Research Group, Konkoly Thege 15-17, H-1121~Budapest, Hungary}
\address[got]{ELTE Gothard Astrophysical Observatory, H-9704 Szombathely, Szent Imre herceg \'ut 112, Hungary}
\address[exoplanet]{MTA-ELTE Exoplanet Research Group, H-9704 Szombathely, Szent Imre herceg \'ut 112, Hungary}
\address[elte]{E\"otv\"os Lor\'and University, Faculty of Science, P\'azm\'any P. st. 1/A, 1171 Budapest, Hungary}
\address[mpe]{Max-Planck-Institut f\"ur extraterrestrische Physik, Giesenbachstrasse, Garching, Germany}
\address[amu]{Astronomical Observatory Institute, Faculty of Physics, A. Mickiewicz University, S\~loneczna 36, 60-286 Pozna\'n, Poland}
\address[gran]{Instituto de Astrof\'isica de Andaluc\'ia (CSIC), Glorieta de la Astronom\'ia s/n, 18008 Granada, Spain}
\address[syd]{Sydney Institute for Astronomy, School of Physics A28, University of Sydney, NSW 2006, Australia}

\begin{abstract}

Here we present the results of visible range light curve observations of ten Centaurs using the \textit{Kepler} Space Telescope in the framework of the K2 mission. Well defined periodic light curves are obtained in six cases allowing us to derive rotational periods, a notable increase in the number of Centaurs with known rotational properties. 

{ The low amplitude light curves of \phlong\, and \kylong\, can be explained either by albedo variegations, binarity or elongated shape. \ydlong\, and \aelong\, could be rotating elongated objects, while \cxshort\, and \vu\, are the most promising binary candidates due to their slow rotations and higher light curve amplitudes.} \vulong\, has the longest rotation period, P\,=\,56.2\,h observed among Centaurs. The P\,$>$\,20\,h rotation periods obtained for the two potential binaries underlines the importance of long, uninterrupted time series photometry of solar system targets that can suitably be performed only from spacecraft, like the \textit{Kepler} in the K2 mission, and the currently running TESS mission.\end{abstract}

\end{frontmatter}

\section{Introduction \label{introduction}}

Centaurs are small solar system objects on non-resonant, giant planet crossing orbits \citep{Gladman2008}, which leads to frequent encounters with the giant planets and results in short dynamical lifetimes. Their origin is the Kuiper belt or the scattering disk, forming a bridge between the transneptunian objects (TNOs) and Jupiter--family comets \citep{tiscareno,disisto,bailey}. Due to their relative proximity they provide an insight into the properties of { outer} solar system objects at the size scale of $\sim$10-100\,km\, \citep{Duffard2014}, which is currently unaccessible in the more distant transneptunian region by typical ground-based observations. Light curve observations and accurate determination of rotational periods of Centaurs are rare. 
In the recent review by \citet{Peixinho2019} there are 20 Centaurs with reliable light curve properties. 

In several cases brightness variations were reported, but no definite periods could be derived, e.g. in the case of 2010\,RF$_{43}$ or 2010\,TY$_{53}$ \citep{Benecchi2013}; or (148975)\,2001\,XA$_{225}$ and (315898)\,2008\,QD$_4$ \citep{Hromakina2018}. This could be due to a common effect of low light curve amplitudes and the faintness of the targets, and/or due to rotation periods longer that could be deduced from typical ground-based observations due to the limited length of the observing blocks. 

As described in \citet{Peixinho2019}   {Centaurs typically rotate faster} than transneptunian objects \citep[mean rotation periods of 8.1\,h and 8.45\,h, respectively,][]{Peixinho2019,Thirouin2019} and they do not show the correlation between light curve amplitude and absolute magnitude observed among transneptunian objects \citep{Duffard2009,Benecchi2013}, which   {might be} explained by the different collisional evolution of small and large transneptunian objects \citep{Davis1997}. 

Periodic light curve variations of a single body can be due to elongated shape or albedo variegations on the surface, or the combination of the two. For minor planets, below the dwarf planet size limit  \citep[radius of $\sim$200--300 km,][]{lineweaver2010}, light curve variation in most cases is explained by shape effects. Binaries can be identified with high probability from light curves only in those special cases when we see a contact binary system under a sufficiently { high} aspect angle, { and binarity is confirmed by multiple epoch observations \citep{Lacerda2007}}. Light curves due to a deformed shape are   {often interpreted} through equlibirium states of a strengthless body ('rubble-pile') in which case the shape is a Jacobi ellipsoid with a well-defined rotation period for a specific density. { This equlibrium density is rather a lower limit for a real object with non-zero internal strength}.  
A rotation period notably shorter than the equlibirium value ({ typically P\,$\gtrsim$\,1\,d, see a detailed discussion {in Sect.~4.2}}) can be interpreted as an indication of a binary system \citep{Leone1984,Thirouin2010,Benecchi2013}. 

There are only two binary Centaurs   {identified} so far: (65489)~Ceto-Phorcys and (42355)~Typhon-Echidna,   {both through direct imaging}. We have to note here that there is some ambiguity in the definition of the Centaurs as   {a} dynamical class. According to the historical definition Centaurs are objects in the giant planet realm whose evolution is currently not controlled by Jupiter \citep[see the discussion in][]{Gladman2008}. A simple   {definition is} that the semi-major axes of their orbits are between that of Jupiter and Neptune. 
The \citet{Gladman2008} dynamical classification scheme   {uses an} additional criterion that a Centaur has to have a perihelion distance q\,$>$\,7.35\,AU and a Tisserand parameter of T$_J$\,$>$\,3.05 to distinguish these objects from Jupiter family comets. 
In this scheme e.g. (60558)~Echeclus (q\,=\,5.8\,AU, T$_J$\,=\,3.03) and (52782)~Okyrhoe (q\,=\,5.8\,AU, T$_J$\,=\,2.95), which are traditionally considered as Centaurs, are classified as 'Jupiter coupled'. The two binaries mentioned above, Ceto-Phorcys and Typhon-Echidna, are classified as Centaurs by the Deep Ecliptic Survey \citep{Elliot2005}, but are considered as scattered disk objects according to \citet{Gladman2008}. We consider them here as these are the only known binaries which are at least dynamically close to the Centaur group, and they are also listed in the recent review of Centaurs by \citet{Peixinho2019}.

\citet{Dotto2008} found a rotation period of 4.43$\pm$0.03\,h and a light curve amplitude of 0.13$\pm$0.02\,mag for Ceto-Phorcys. This is an unexpected result, as according to \citet{Grundy2007} this is a tidally evolved and spin locked binary system, with a small orbital eccentricity (e\,$\leq$0.013) and orbital period of P\,=\,9.554\,d. 
The Typhon-Echidna system is the other known binary Centaur, discovered by \citet{Noll2006}, with an orbital period of P$_{orb}$\,=\,18\fd98 and semi-major axis of $a$\,=\,1629\,km \citep{Grundy2008}. In contrast to Ceto-Phorcys the binary orbit of Typhon-Echidna is rather eccentric (e\,=\,0.53), showing that this is not a tidally evolved system. \citet{Thirouin2010} obtained a tentative rotation period of P$_{rot}$\,=\,9.67\,h and a small light curve amplitude of $\Delta m$\,=\,0\fm07$\pm$0\fm01, consistent with other studies reporting on nearly flat light curves \citep{ortiz2003,sheppard2003}.

A large fraction of binaries originally in high eccentricity orbits can evolve to circular and very tight orbits due to Kozai effects \citep{Porter2012}. An originally triple system that loses a component will also end up in a very tight pair \citep{Margot2015}.  The angular resolution of the Hubble Space Telescope -- that   {has} detected most of the known transneptunian and Centaur binaries -- allows the detection of a nearly equal brightness binary with a semi-major axis of $\sim$400\,km at 10\,AU (typical perihelion distance of Centaurs)  {; proportionally} wider systems can be discovered at larger heliocentric distances. 
Due to the lack of suitable spatial resolution more compact systems can typically be discovered through the detection of their characteristic light curves, { i.e. large, $\Delta m$\,$\gtrsim$\,1\,mag amplitudes with U-shaped maxima and V-shaped minima} \citep[see e.g.][for a discussion of contact binary systems in the plutino population]{Thirouin2018}. In some rare cases binary nature can be deduced from stellar occultation observations, as in the case of 2014\,MU$_{69}$ \citep{Moore2018}.

As suggested by \citet{Thirouin2018}, nearly half of the plutino population may be contact or   {a tight binary system}. Since plutinos are thought to be one of the parent populations of Centaurs \citep{disisto2010}, one can expect a similar abundance of contact and tight binaries in the Centaur population, too, assuming that tight systems remains intact in giant planet encounters. 

Studies of a large sample of minor planet light curves observed in the framework of the K2 mission of the \textit{Kepler} Space Telescope \citep{howell2014} showed an overabundance of long (up to several days) rotation periods among main-belt asteroids \citep{Szabo2016,Molnar2018}. In the case of Jovian Trojans a { binary} fraction of 6--36\% \citep{ryan} and $\sim$20\% \citep{Szabo2017} was estimated from the data.  These studies were carried out in the course of systematic programs in the K2 mission, aimed at obtaining light curves of solar system targets, including main belt asteroids 
\citep{Szabo2015,Szabo2016,Berthier2016,Molnar2018}, Jovian Trojans \citep{ryan,Szabo2017}, transneptunian objects
\citep{Pal2015,Pal2016,Benecchi2018} and irregular moons of giant planets 
\citep{Kiss2016,Farkas2017}. These observations provided continuous light curves which had significantly longer time-spans (up to 80 days) than ground-based measurements and therefore could break the ambiguity of rotational periods caused by daily aliases. 

In this paper we report on observations of ten Centaurs: \kylong, \ydlong, \gxlong, \jjlong, \pllong, \vulong, \phlong, \fzlong, \aelong, and \cxlong\footnote{We use the provisional designations to identify our targets throughout the paper}, observed with \textit{Kepler} in the K2 campaigns. One target, \kylong, has previous light curve measurements, and for this target we provide an updated, more accurate rotation period and light curve. The other nine Centaurs have no light curve measurements reported in the literature.   {Due to poorly constrained light curve properties for four targets we add reliable measurements for five objects to the group of Centaurs with known rotation periods and light curves.} We also perform simple calculations to deduce whether the light curve variation of our targets can be caused more likely by shape effects or binarity.


\section{Observations, data reduction and photometry}

\begin{table*}
\caption[]{Summary of K2 light curve observations of our Centaur sample. The columns are: 
Name -- provisional designation of the target; 
Cam. -- K2 campaign number; 
Start -- Start date of the K2 observations (Julian date);
End -- End date of the K2 observations (Julian date); 
Length -- total length of the observations (day);
Duty cycle -- fraction of frames used for light curve photometry;
r$_h$, $\Delta$ and $\alpha$ -- heliocentric distance,  target to observer distance and phase angle range during K2 measurements;
m$_{11}^\mathrm{R}$ and m$_{110}^\mathrm{R}$ -- phase angle uncorrected and corrected USNO B1.0 R-band absolute magnitude of the targets, derived from our K2 observations.}
\scriptsize\addtolength{\tabcolsep}{-3pt}
\begin{tabular}{cccccccrrrrr}
\hline
Name  &  Cam.  &  Start  &  End  &  Length  &  \#frame  &  Duty  &  
r$_h$ & $\Delta$ & $\alpha$ &
m$_{11}^\mathrm{R}$ &  m$_{110}^\mathrm{R}$ \\
&& (JD) & (JD) & (day) & & cycle & (AU) & (AU) & (deg) & (mag) & (mag) \\
\hline

\ky & C04  &  2457061.7951 & 2457090.4838 & 28.689 & 1302 & 0.928 &10.729...10.774  &   9.931...10.358   &  3.129...4.837 & 9.85$\pm$0.06   & 9.43$\pm$0.10  \\
\yd & C16 &  2458131.0852 & 2458150.7219 & 19.637 & 867 & 0.902  & 14.869...14.906  &  14.474...14.762   &  3.429...3.788 & \multirow{2}{*}{10.13$\pm$0.27} & \multirow{2}{*}{9.75$\pm$0.27} \\
\yd & C18 &  2458263.4745 & 2458302.3800 & 38.905 & 1401 & 0.734  & 15.125...15.204 & 14.499...15.169 & 3.086...3.909 &  &  \\
\gx  &  C11  & 2457669.6537 & 2457679.5845 & 9.931 & 300 & 0.617 & 16.787...16.791  &  16.349...16.501   &  3.230...3.400 & 8.48$\pm$1.55 & 8.14$\pm$1.56\\
\jj  &  C11 & 2457682.7721 & 2457692.8663 & 10.094 & 395 & 0.800  & 23.994...24.000  &  23.644...23.814   &  2.320...2.405 &  7.06$\pm$0.75 & 6.81$\pm$0.75  \\

\plshort  & C12  &  2457754.3510 & 2457799.6318 & 45.281 & 1148 & 0.518 & 21.553...21.616  &  21.042...21.732   &  2.151...2.591 & 8.25$\pm$0.25 &  7.99$\pm$0.25 \\
\vu  & C13 & 2457850.1437 & 2457880.8963 & 30.753 & 1022 & 0.679 & 15.583...15.588  &  15.073...15.577   &  3.148...3.594 & 8.39$\pm$0.44	&	8.16$\pm$0.45\\
\ph  & C12 & 2457756.1287 & 2457786.1048 & 29.976 & 859 & 0.586 &	24.735...24.767  &  24.334...24.813   &  2.078...2.329 & 9.53$\pm$0.21	&	9.17$\pm$0.22\\
\fzshort & C15 & 2458179.5332 & 2458246.4125 & 66.879 & 1108 & 0.338 &  14.694...14.781 & 13.997...14.956 & 2.507...3.823 & 10.66$\pm$0.47 & 10.24$\pm$0.47 \\
\aeshort & C16 & 2458122.4827 & 2458151.1715 & 28.689 & 1245 & 0.887 & 16.977...16.996  &  16.508...16.965 &  2.907...3.339 &  8.64$\pm$0.17	&	8.31$\pm$0.17 \\
\cxshort & C18 & 2458288.2605 & 2458298.2321 & 9.972 & 379 & 0.773  & 10.675...10.685  & 10.377...10.549  & 5.366...5.537 & 11.33$\pm$0.30 & 10.68$\pm$0.30 \\

\hline
\end{tabular}
\label{table:k2obs}
\end{table*}

\begin{figure*}
\includegraphics[width=\textwidth]{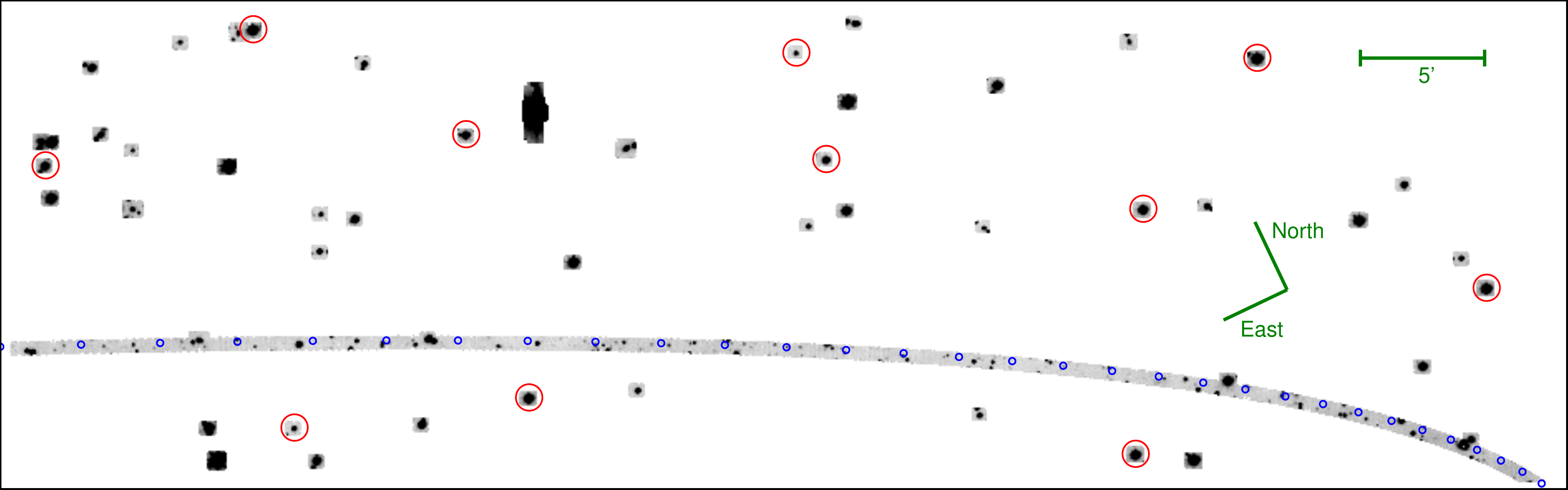}
\caption[]{The field-of-view of \textit{Kepler} in which \ky{} was moving during the K2 Campaign 4, as an example. The stars involved in the determination of the absolute and differential astrometric solutions are indicated by red circles. The small blue circles indicate the position of the targets with a 1-day stepsize throughout the observations. The field is shown in the CCD frame, therefore the image is rotated with respect to the standard view.}
\label{fig:fields}
\end{figure*}

\textit{Kepler} observed numerous Solar System objects during the K2 mission. The observing strategy and data reduction steps of centaurs have been analogous to other TNO and asteroid targets that we   {previously} published \citep{Pal2015,Kiss2016,Molnar2018}. Since \textit{Kepler} observed only selected pixels during each 60-80\,d long Campaign, pixels over $\sim 30$\,d long arcs of the apparent trajectories of the target were allocated for each Centaur (see Fig.~\ref{fig:fields} for an example). 

We processed the \textit{Kepler} observations with the \textit{fitsh} software package \citep{Pal2012}. First, we assembled the individual Target Pixel Files of both the track of the target and that of the nearby stars into mosaic images.   {Astrometric solutions were derived for every mosaic frame in the campaign, using the Full Frame Images (acquired once per campaign) as initial hints, and the individual frames were registered into the same reference system.} We then enlarged the images by roughly 3 times, and transformed them into RA-Dec directions. This enlargement helped to decrease the fringing of the residual images in the next step, where we subtracted a median image from all frames. The median was created from a subset of frames that did not contain the target. We applied simple aperture photometry to the differential images based on the ephemeris provided by the JPL HORIZONS service. 

We then discarded data points that were contaminated by the residuals of the stellar images,   {saturated} columns and crosstalk patterns from the camera electronics 
 -- this is characterised by the duty cycle, the ratio of the number of frames used for the final light curve derivation and the total number of frames on which the target was theoretically visible. While this ratio is typically well above 50\%, it is only $\sim$34\% for \fz, which was very faint, and thus very sensitive to any structures in the background. We had to discard a large number of frames that were affected by stellar residuals, crosstalk patterns, or   {in which} the object was not detected.

The light curves obtained were analysed with a residual minimization algorithm \citep{Pal2016,Molnar2018}. In this method we fit the data with a function $A + B \cos(2\pi f \Delta t) + C \sin(2\pi f \Delta t)$, where $f$ is the trial frequency, $\Delta t = T - t$,  T the approximate center of the time series, and A, B, and C are parameters to the determined. We search for the minimum in the dispersion curves for each frequency. As demonstrated in \citet{Molnar2018} the best fit frequencies obtained with this method are identical to the results of Lomb–Scargle periodogram or fast Fourier transform analyses, with a notably smaller general uncertainty in the residuals. 


\section{Results \label{sect:results}}

\subsection{Absolute magnitudes}

We determined the absolute magnitudes of the targets, transformed from the K2 observations to the USNO B1.0 R-band photometric system \citep{Monet2003}, in the same way as in \citet{Pal2015}. We calculated both the phase angle uncorrected (\mruncorr) and phase angle corrected (\mrcorr) {absolute magnitudes}. As the phase angle ranges of the observations were rather small (a maximum of 1\fdg7 in the case of \ky) we could not fit a phase angle correction curve when calculating \mrcorr, but used a  $\beta_R$\,=\,0.104$\pm$0.074\,mag\,deg$^{-1}$ linear phase angle correction, obtained as the median values of the R-band correction factors of Centaurs in \citet{ayala}. While there is a specific phase correction coefficient determined for \ky, it is based on sparse and uncertain data \citep{AC2016}, therefore we used the coefficient above even in this case. These absolute magnitudes are listed in Table~\ref{table:k2obs} along with the basic observational parameters, using data from the Minor Planet Center. 

\subsection{Light curves \label{sect:lightcurves}}

We were able to determine light curve periods from the periodograms for six targets. The folded light curves and residual dispersion plots are presented in Fig.~\ref{fig:k2lcs}, and the rotation periods and light curve amplitudes observed are summarized in Table~\ref{table:k2periods}.  

In all cases we accepted the most prominent peak in the phase dispersion versus frequency plot as the actual primary light curve period. The quality of the light curve frequency/period determination is characterised by the accuracy of the frequency determination (converted to period uncertainty (h) in Table~\ref{table:k2periods}), and also by the ratio of the light curve peak over the r.m.s. of the phase dispersion at other frequencies. This latter was calculated for the whole frequency range investigated (S$_{\mathrm f0}$ in Table~\ref{table:k2periods}) and also for a narrower frequency range of $\pm$1\,d$^{-1}$ around the peak (S$_{\mathrm fp}$ in Table~\ref{table:k2periods}). 

Period and amplitude uncertainties were also calculated with the {\sl FAMIAS} and {\sl Period04} methods using Fourier-transforms \citep{Lenz2004,Zima2008}. We calculated the formal least-squares uncertainties with both softwares, plus the Monte Carlo module of Period04 that generates sets of artificial data with randomized noise based on the residual light curve, and tries to fit them with the input frequency set. The latter method gave elevated uncertainties in three cases, most notably for \vu\, and \ph, due to the low signal-to-noise ratio of the fitted frequency and the large scatter of data points. We also calculated the relative uncertainties for the main frequency components that generally agreed with the error estimates for the full amplitudes. We chose the larger of these estimates for { each} target (see the period and amplitude uncertainties in Table~\ref{table:k2periods}). 

To characterise the possible double peak nature we folded the light curves with the double peak period and calculated the significance S$_{\mathrm dp}$ of the difference between the two light curve halfs (phases 0\,$\leq$\,$\phi$\,$<$\,0.5 and 0.5\,$\leq$\,$\phi$\,$<$\,1) following \citet[see eqs. 2, 3 and 4]{Pal2016}. These significance values were calculated for a series of bin numbers N\,=\,16...24 which resulted in only slightly different S$_{\mathrm dp}$ values for the same target. The actual mean S$_{\mathrm dp}$-s are listed in Table~\ref{table:k2periods}.
Following \citet{Pal2016} we used the criterion that for a detectable double peak behaviour S$_{\mathrm dp}$\,$\geq$\,3. In our sample only \aeshort\, has S$_{\mathrm dp}$\,=\,1.8, for all other targets S$_{\mathrm dp}$\,$>$\,3, indicating that a double peak light curve is likely in these latter cases.

\subsection{Discussion of the individual targets}

{ \kylong} was discovered in 2002 by   {Trujillo, C. A. \& Brown, M. E..} \citet{Thirouin2010} reported on a single peak rotational period of 3.56\,h or 4.2$\pm$0.05\,h with an amplitude of 0.13$\pm$0.01\,mag. \citet{Duffard2014} modeled the thermal emission of \ky{} using Herschel/PACS measurements of the "TNOs are Cool!" Open Time Key Program, using a NEATM model with fixed beaming parameter of $\eta$\,=\,1.2, and obtained an effective diameter and albedo solution of D\,=\,47$^{+3}_{-4}$\,km and p$_V$\,=\,5.7$^{+1.1}_{-0.7}$\%. Our new rotation period is P\,=\,8.4996$\pm$0.0036\,h, with an asymmetric, double peak light curve. The amplitude of the first maximum is $\Delta m_1$\,=\,0.090$\pm$0.009\,mag, with a secondary maximum of $\Delta m_2$\,=\,0.028$\pm$0.008\,mag, i.e. the first peak is taller by 0.062\,mag. The new spin period is the double period of the 4.2\,h found by \citet{Thirouin2010}.

{ \ydlong} was observed in the K2 missions in two campaigns, C16 and C18 (see also Table~\ref{table:k2obs}). A well defined, double peak light curve is obtained with a period of P\,=\,10.1590$\pm$0.0008\,h, and two similar light curve amplitude peaks of 
$\Delta m$\,=\,0.202$\pm$0.028\,mag and $\Delta m$\,=\,0.180$\pm$0.034\,mag. 

For { \vulong} we obtained a light curve with a period of P\,=\,28.12$\pm$1.66\,h and light curve amplitude of $\Delta$m\,=\,0.38$\pm$0.05\,mag, assuming a single peak light curve. If the double peak light curve is considered (P\,=\,56.2\,h), it is the Centaur with the longest rotation period ever observed. { The double peak period seems to be more likely due to the different first and second peaks observed in the double peak light curve (S$_{\mathrm dp}$\,=\,3.2)}

\begin{figure*}
\includegraphics[width=\textwidth]{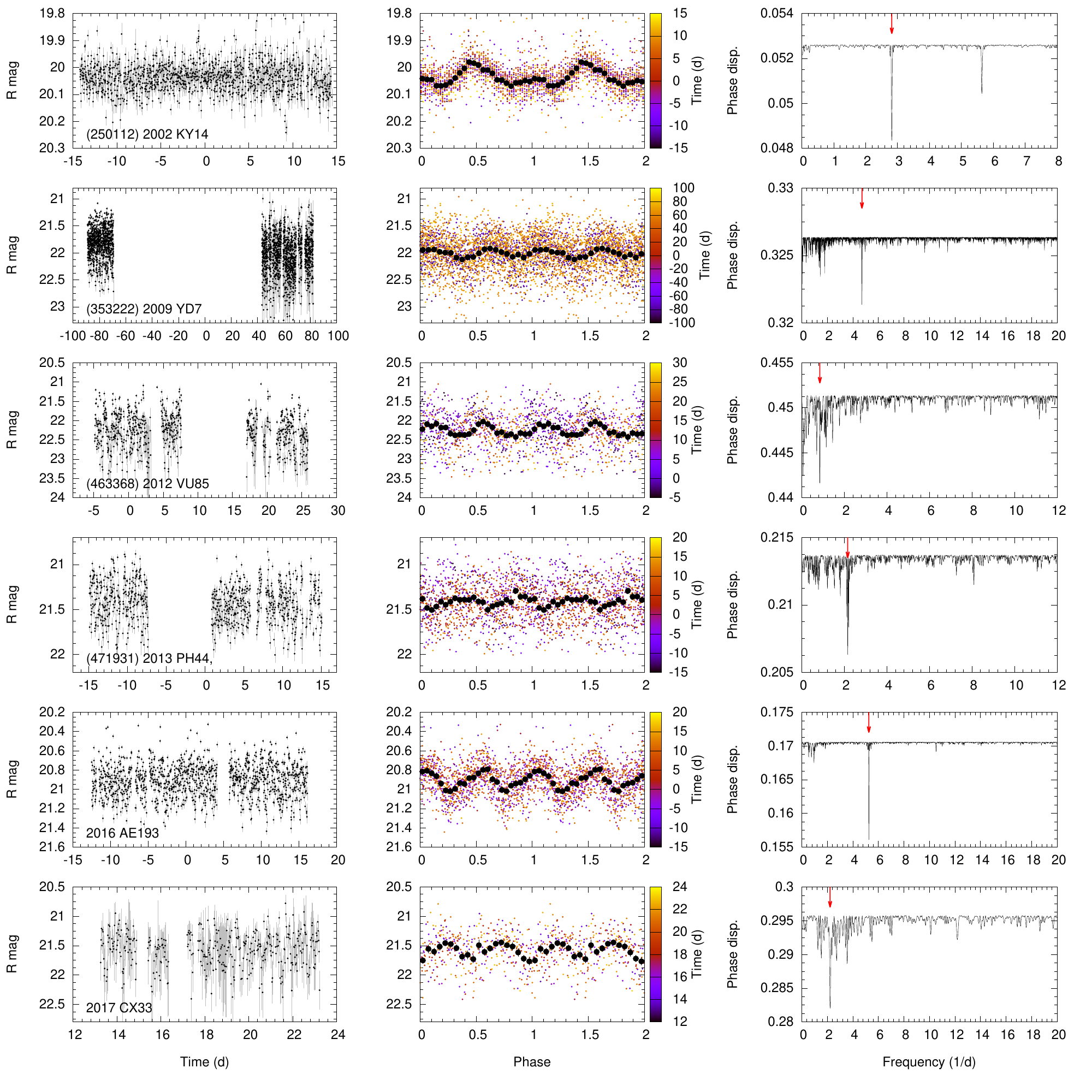} 
\caption[]{The observed light curve (left), the phase curve (middle), in all cases folded with the most probable period (middle), and the residual dispersion versus frequency plot (left). 
In the middle column the color scale represents dates, BJD-t$_0$, as indicated at the side of the figures. In the normalised residual plots red arrows mark the primary periods detected \label{fig:k2lcs}}
\end{figure*}

A single peak light curve of { \phlong} is detected with a period of P\,=\,11.08$\pm$0.12\,h, and light curve amplitude of $\Delta m$\,=\,0.15$\pm$0.04\,mag. With the corresponding P\,=\,22.16$\pm$0.24\,h double peak period the light curve is notably asymmetric, as presented in Fig.~\ref{fig:k2lcs}, making the double peak period more likely.

{ \aelong} has a single peak light curve with a period of P\,=\,4.556$\pm$0.013\,h, and light curve amplitude of $\Delta m$\,=\,0.228$\pm$0.014\,mag. { The double period light curve shows no significant asymmetry, and the light curve asymmetry parameter derived (S$_{\mathrm dp}$\,=\,1.8, see above)} indicates that the light curve is { rather} single peak (a light curve folded with the double peak period is presented in Fig.~\ref{fig:k2lcs} for consistency). { However, this does not exclude that the double peak period is the rotation period of this target. E.g. a sufficiently symmetric triaxial ellipsoid -- often used in simple asteroid shape modelling -- produces a light curve with two identical half periods. }

{ \cxlong} is moving on a very high inclination orbit (see Table~\ref{table:k2periods}). It has a rotation period of P\,=\,21.51$\pm$0.13\,h (double peak) with a light curve amplitude of $\Delta m$\,=\,0.27$\pm$0.11\,mag. 

\begin{figure*}
\includegraphics[width=\textwidth]{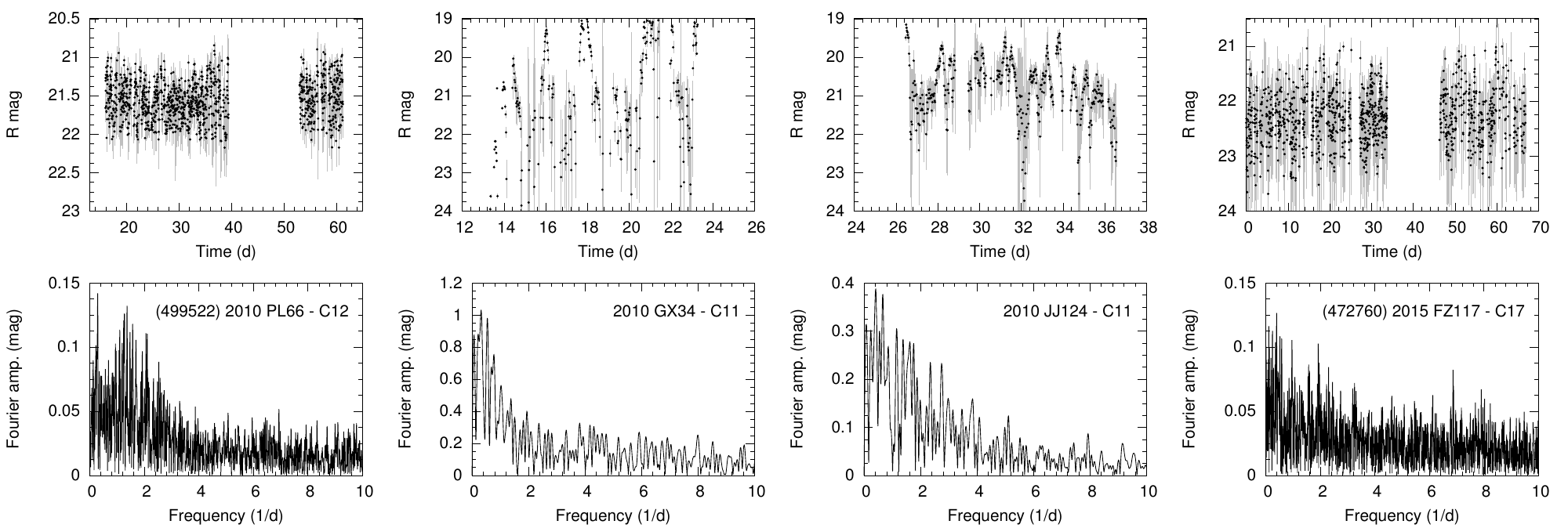} 
\caption[]{The observed light curve (top row) and the corresponding Fourier amplitude plots (bottom row) of those three Centaurs (\plshort, \gx\, and \jj) for which no rotation period could be obtained. {Fourier amplitude plots are presented here instead of the dispersion residual plots as these were used to estimate the amplitude upper limits in the case of targets with no light curve period detected.}}
\label{fig:nondetect}
\end{figure*}

For four of our targets no unambiguous rotation period could be obtained (\pllong, \gxlong, \jjlong, \fzlong, see Fig.~\ref{fig:nondetect}). For these objects we provide an upper limit on the light curve amplitude only (see Table~\ref{table:k2periods}). As seen in Fig.~\ref{fig:nondetect} the Fourier amplitude depends strongly on the spin rate -- there is a significant increase towards smaller frequencies / longer light curve periods, i.e. a light curve could have been detected at higher frequencies with a smaller amplitude, and we more likely miss light curve periods if P\,$\geq$\,1\,d for these targets.   

Altogether we add one updated light curve (\ky), and five new ones to the list of Centaurs with known light curve properties, previously containing 20 targets \citep{Peixinho2019}.

\begin{figure}
\includegraphics[width=\textwidth]{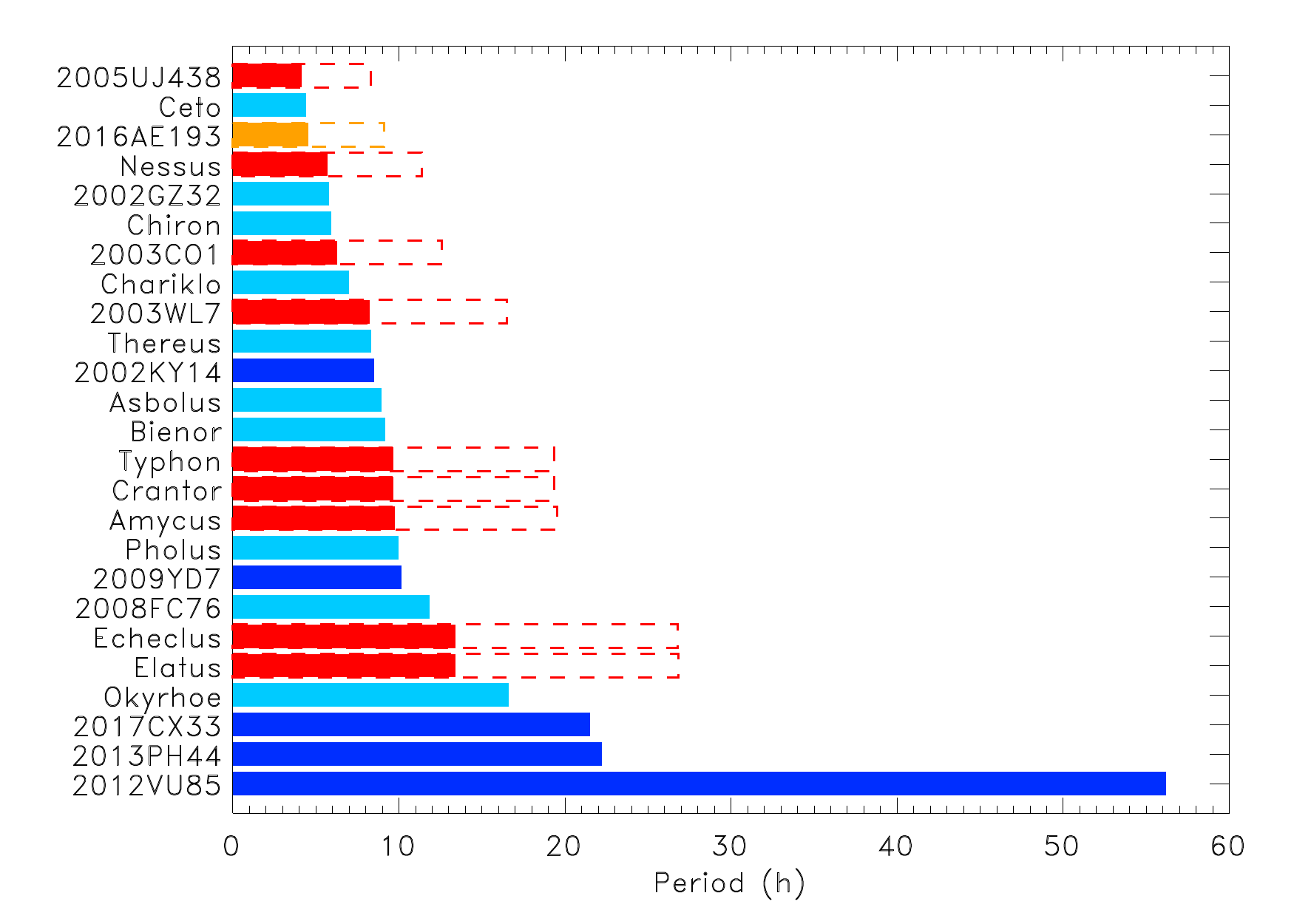}
\caption{Likely rotation periods of Centaurs in the literature and in our present work, sorted by increasing rotation period. The bar colours correspond to: 
red -- literature, single peak period;
orange -- this study, single peak period;
light blue -- literature, double peak period;
dark blue --- this study, double peak period. In the case of single peak periods we also   {include} the double peak periods with dashed lines.  \label{fig:periodplot}}
\end{figure}

\begin{table*}
\caption[]{Summary of the derived rotation periods and amplitudes. $\langle R \rangle$ is the mean USNO B1.0 R-band brightness of the target. We list both the single peak and double peak periods (P$_s$ and P$_d$, respectively), the one marked with bold-face characters is the more likely one according to our criteria. S$_{\mathrm f0}$ and S$_{\mathrm fp}$ are the significances of the light curve period determination, and  S$_{\mathrm dp}$ is the significance parameter describing the possible double peak behaviour (see text for details).
The amplitude upper limits are determined for possible periods shorter than 24\,h. The last four columns list the main orbital parameters.}
\scriptsize
\begin{tabular}{cccccccccccccc}
\hline
Name  & $\langle R \rangle$  & P$_s$ & P$_d$ & $\Delta$m  & $\delta\Delta$m & 
S$_{\mathrm f0}$ & S$_{\mathrm fp}$ & S$_{\mathrm dp}$ & a & q & e & i \\
& (mag) & (h) & (h) & (mag) & (mag) & & & & [au] & [au] & & [$^\circ$]\\
\hline
\ky  &  20.04 & -- & { 8.4996$\pm$0.0036} &  0.0899 & 0.006   & 16.6 & 9.6 & 9.4 & 12.43 & 8.60 & 0.313 & 19.5 \\
\yd  &  21.82 & 5.0795$\pm$0.0004 & { 10.1590$\pm$0.0008} & 0.21 & 0.020  & 20.7 & 14.2 & 4.2 & 121.96& 13.38& 0.890 & 30.81\\
\gx  &  20.69 &  -- &  -- & $<$0.6 & --            & & & -- & 29.01 & 16.57& 0.429 & 11.54\\
\jj  &  20.83 &  -- &  -- & $<$0.5 & --            & & & -- & 85.55 & 23.61& 0.724 & 37.70\\
\plshort  &  21.56 &  -- &  -- & $<$0.2 & --       & & & -- & 21.12 & 13.08& 0.381 & 24.35\\
\vu  &  22.31 & 28.12$\pm$1.66 & { 56.24$\pm$3.32} & 0.38 & 0.05 & 11.2 & 4.8 & 3.2 & 29.15 & 20.10& 0.311 & 15.10\\
\ph &  21.42  & 11.08$\pm$0.12 & { 22.16$\pm$0.24} & 0.15 & 0.04  & 14.4 & 7.8 & 3.6 & 19.63 & 15.53& 0.209 & 33.53\\
\aeshort &  20.91 & { 4.556$\pm$0.013} & 9.112$\pm$0.026 & 0.228 & 0.014 & 27.6 & 9.6 & 1.8 & 30.40 & 16.52 & 0.467 & 10.27 \\
\fzshort & 22.31 & -- & -- &  $<$0.2 & -           & -- & -- & -- & 22.99 & 13.15 & 0.428 & 6.81 \\
\cxshort & 21.53 & 10.755$\pm$0.064 & { 21.51$\pm$0.13} & 0.27 & 0.11      & 11.1 & 5.4 & 3.3 & 73.48 & 10.45 & 0.858 & 72.05 \\
\hline
\end{tabular}
\label{table:k2periods}
\end{table*}

\subsection{Comparison with Centaurs with known light curves}

The most recent review by \citet{Peixinho2019} lists light curve periods for 20 Centaurs including \ky\,($\equiv$\,2007\,UL$_{126}$), and we have considered this sample as a reference sample for a comparison with our targets. The targets included are: 
(2060) Chiron,
(5145) Pholus,
(7066) Nessus, 
(8405) Asbolus,
(10199) Chariklo,
(31824) Elatus,
(32532) Thereus,
(42355) Typhon,
(52872) Okyrhoe,
(55567) Amycus,
(60558) Echeclus,
(65489) Ceto,
(83982) Crantor,
(95626) 2002\,GZ$_{32}$,
(120061) 2003\,CO$_1$,
(136204) 2003\,WL$_7$,
(145486) 2005\,UJ$_{438}$,
(281371) 2008\,FC$_{76}$. 
We used the preferred single peak or double peak light curve periods as rotation periods as given in \citet{Peixinho2019}, counter checked with the original papers \citep[see the references in][]{Peixinho2019}. Note that the period of 8.32\,h for 2005\,UJ$_{438}$ is the double peak period according to \citet{Thirouin2010}, and that for Ceto and Typhon the light curves periods used here are not the binary orbital periods, as discussed in Sect.~1. 

We compare the rotation periods of the Centaurs in the reference sample with our targets in Fig.~\ref{fig:periodplot}. Our targets are presented in this plot with their preferred single or double peak periods, as discussed in Sect.~\ref{sect:lightcurves} above.

\begin{figure}
\includegraphics[width=0.75\textwidth]{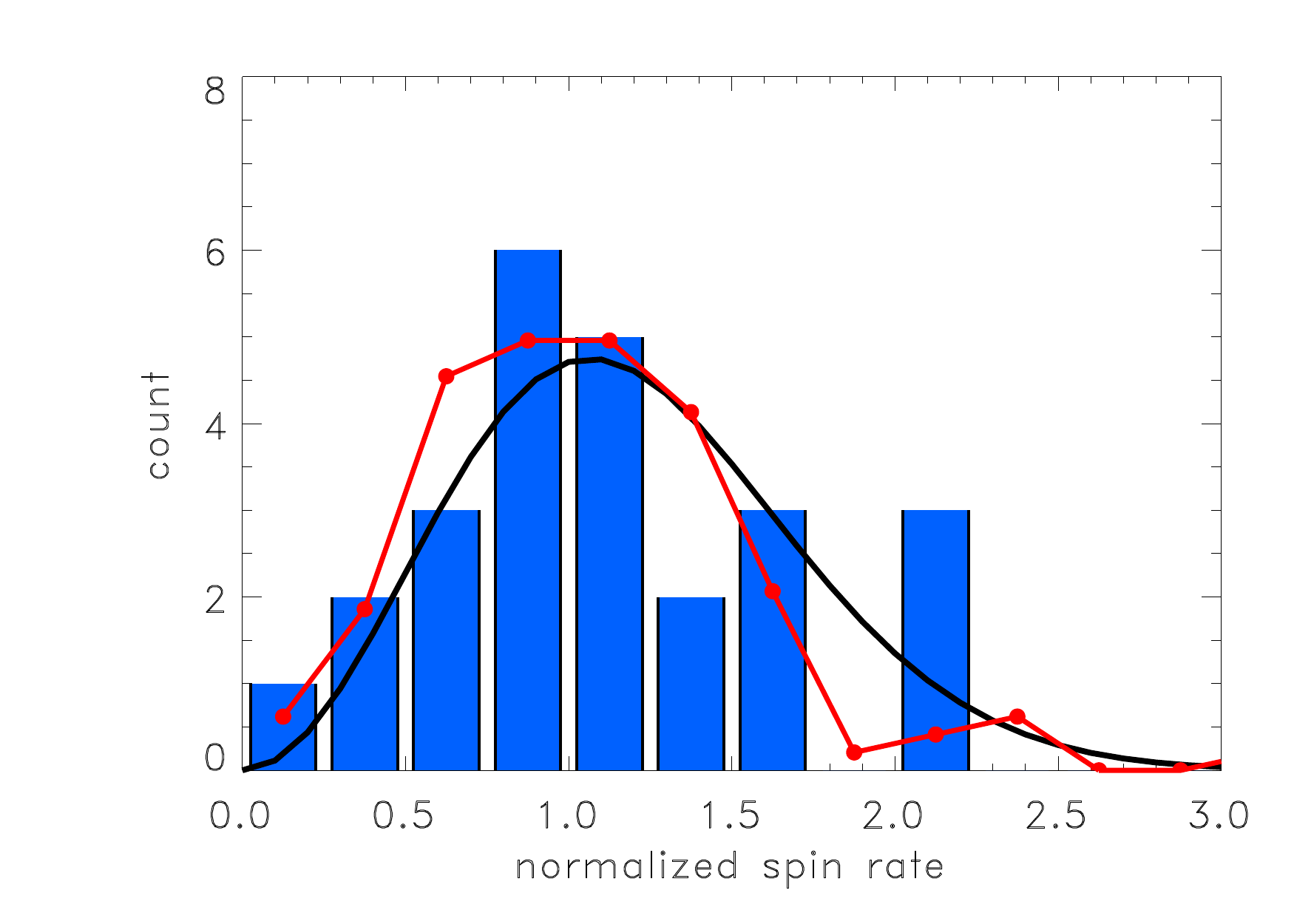}
\caption{Histogram presenting the spin rate distribution of Centaurs with our targets included, as a function of normalised spin rate (blue bars). The red curve represents the spin rate distribution of transneptunian objects as obtained from the LCDB, normalised to the total number of Centaurs in our sample. The black solid curve shows the Maxwellian fit to the Centaur data.  \label{fig:rothist}}
\end{figure}

Using the rotation periods as presented in Fig.~\ref{fig:periodplot}   {the} Centaurs with the three longest rotation periods are from our sample (\cxshort, \ph, \vu). When considering double peak periods for all targets, however, there are several other objects with similar rotation periods (Typhon, Crantor, Amycus, Echeclus, Elatus). The single important feature is the quite long, P\,=\,56.20\,h rotation period of \vu, not seen previously among Centaurs. Using the whole Centaur sample the mean rotation period is $\langle P \rangle$\,=\,9.2\,h (8.9\,h without our targets), which is now between the mean rotation period of the cold classicals (9.48$\pm$1.53\,h) and the rest of the TNOs (8.45$\pm$0.58\,h), as obtained by \citet{Thirouin2019}. The TNO sample in the Light Curve Database \citep[LCDB][]{Warner2009} has a spin rate distribution rather similar to that of Centaurs (red curve in Fig.~\ref{fig:rothist}), with a median rotation period of $\langle P \rangle$\,=\,8.84\,h. A Maxwellian fit to the spin rate distribtuion \citep[see e.g.][for a discussion]{Pravec2000} seems to be an acceptable model as it provides a reduced-$\chi^2$ value of $\lesssim$\,1, using the square root of the number of objects in the specific bins as uncertainties.  

\section{Rotating elongated bodies versus binarity} 

\subsection{Density estimates from Jacobi ellipsoid models}

\begin{figure}
\includegraphics[width=0.75\textwidth]{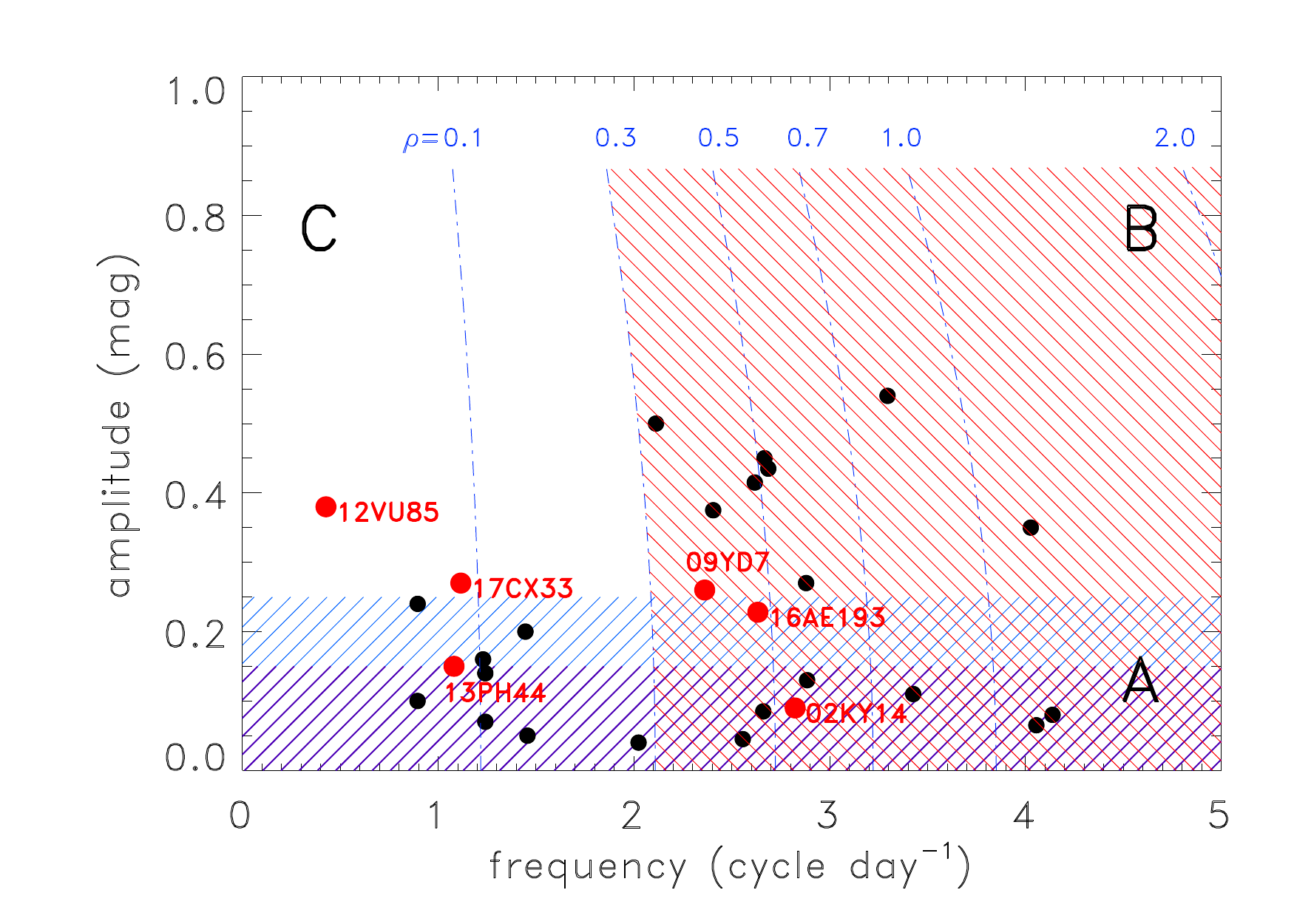}
\caption{Light curve amplitude versus frequency of the reference sample Centaurs (black dots) and our six targets (red dots). Blue dash-dotted curves represent the rotational frequencies and light curve amplitudes corresponding to the rotation of a strengthless body, modeled as Jacobi ellipsoids, of a specific density. The densities of the curves are shown at the top in g\,cm$^{-3}$ units. In the { blue and purple} shaded areas (below $\delta$m\,$\leq$\,0.25 or 0.15\,mag, region~A) light curves can be explained either by albedo variegations, deformed shape or binarity. { Targets in the red shaded area (region~B), right of the  $\rho$\,=\,0.3\,g\,cm$^{-3}$ curve, could be elongated due to rotation. Objects in region~C should have densities notably below $\rho$\,=\,0.3\,g\,cm$^{-3}$ in order to be elongated from rotation, and can be considered as binary candidates (see the text for a detailed discussion).  }
 \label{fig:freqampl}}
\end{figure}

\citet{Leone1984} and \citet{Sheppard2004} identified three main zones on the light curve amplitude versus rotational frequency plane (see Fig.~\ref{fig:freqampl}), re-evaluated by \citet{Thirouin2010} and \citet{Benecchi2013}. Light curve variations of objects with small amplitudes ($\Delta m$\,$\leq$\,0.25\,mag or 0.15\,mag) can either be caused by albedo and shape features or can as well be binaries ({ region~A}  in Fig.~\ref{fig:freqampl}). If the rotational equlibrium of a strengthless body is considered and approximated by a Jacobi ellipsoid, constant density curves can be drawn (blue dash-dotted curves in Fig.~\ref{fig:freqampl}). We list the densities estimated this way for our targets in the last column of Table~\ref{table:tidal} following \citet[see eqs. 1 \& 2 in][and references therein]{Lacerda2007}, assuming $\vartheta$\,=\,$\pi/2$ aspect angle, { i.e. equator-on viewing geometry and maximum light curve amplitude}. Objects to the right of a curve of a constant density (e.g. 0.3\,g\,cm$^{-3}$ for Centaurs, region~B) are likely rotating single bodies, if their rotational speed in below the breakup limit (4.0\,cycle\,day$^{-1}$ for 0.3\,g\,cm$^{-3}$). The rotation of the objects to the left is too slow to cause elongation and a corresponding rotational light curve. For these objects the light curves are often explained by binarity \citep[e.g.][]{Leone1984,Sheppard2004}. 

For three of our targets the estimated Jacobi ellipsoid densities are in the order of $\sim$0.5\,g\,cm$^{-3}$ ($\rho_{\mathrm JE}$\,=\,0.54, 0.39 and 0.48\,g\,cm$^{-3}$ for \ky, \yd\, and \aeshort, respectively, see Table~\ref{table:tidal}), inside the range expected for smaller (D\,$<$\,500\,km) transneptunian objects and Centaurs \citep{Grundy2019,Kiss2019}. The Jacobi ellipsoid density estimates are, on the other hand, notably lower for \vu, \cxshort\, and \ph\, ($\rho_{\mathrm JE}$\,$<$\,0.1\,g\,cm$^{-3}$), outside the range of densities plausibly considered, indicating that the light curves in these cases cannot be explained by equlibirium figures of rotating strengthless bodies. As discussed above, objects in this part of the light curve amplitude vs. rotational frequency plot may be considered as binaries. However, due to the low amplitude ($\Delta m$\,$\leq$\,0.15\,mag) the light curves of \ky\, and \ph\, may as well be explained by albedo variegations on the surface, in addition to possible binarity or elongated shape. Also, the light curve amplitude of \aeshort\, is below the 0.25\,mag limit originally considered for surface variegations. 

\subsection{Characterisation of potential binarity}

It is a question in the case of a binary whether the observed rotation period is the common, synchronised period of a binary,   {or if we can see the light curve} of a single body, rotating with an angular speed different from the orbital one. In the main belt small binary asteroids (D\,$\lesssim$10\,km) are typically asynchronous if their rotation period is P\,$\lesssim$\,8\,h \citep{Pravec2007}. Synchronous binaries are found for P\,$\geq$\,8\,h, usually at the D\,$\approx$\,10\,km sizes, but there are synchronous systems with D\,$\approx$\,100\,km as well ((90)~Antiope and (617)~Patroclus-Menoetius), bracketing the size range of the Centaurs in our sample. Large asteroids (D\,$\geq$\,100\,km) with small satellites also  {typically rotate} faster {\citep[P\,$\lesssim$\,8\,h][]{Pravec2007}}. 

In the plutino population, a likely parent population of Centaurs, \citet{Thirouin2018} estimated that the incidence rate of contact binaries   {could} be as high as $\sim$50\%.   {In the transneptunian region there is an overabundance of nearly equal-brightness (and therefore probably nearly equal-mass) binaries among the resolvable systems \citep{Noll2008}, and a large fraction, even close to 100\% among cold classical Kuiper belt objects \citep{Noll2014,Fraser2017}.}

While we cannot unambiguously identify a binary system from the light curve and rotation period alone, a simple check can be performed to show whether a specific system could potentially be a binary based on its light curve period and absolute magnitude. To characterise a system in this way we use the estimated 'separation', $a_{bin}$, the semi-major axis of the orbit of the potential binary. We assume that the binary has two equally sized and equal mass components \citep{Noll2008}. In the case of our Centaur reference sample we used the radiometric size estimates based either on Herchel/PACS \citep{Duffard2014,Fornasier2013} or WISE \citep{Mainzer2016} observations, whenever these were available; when radiometric size was not available we simply used the default size (or albedo and absolute magnitude) estimate in \citet{Peixinho2019}, and used this value to calculate the binary diameters and volumes \citep[see e.g.][]{Vilenius2014}. The binary separation, $a_{bin}$ is obtained from Kepler's third law, assuming a density of 0.7\,g\,cm$^{-3}$ to obtain the mass, characteristic for 10-100\,km-sized Kuiper belt bodies and Centaurs \citep[see e.g.][for a latest compilation of Kuiper belt densities]{Grundy2019,Kiss2019}. The densities estimated for Ceto-Phorcys \citep[$\rho$\,=\,1.37$^{+0.66}_{-0.32}$\,g\,cm$^{-3}$][]{Grundy2007} and Typhon-Echinda \citep[$\rho$\,=\,0.44$^{+0.44}_{-0.17}$\,g\,cm$^{-3}$][]{Grundy2008} are at the lower/upper extremes of the densities of $\sim$100\,km-sized objects, and therefore may not be representative for the whole population.

\begin{figure}
\includegraphics[width=0.75\textwidth]{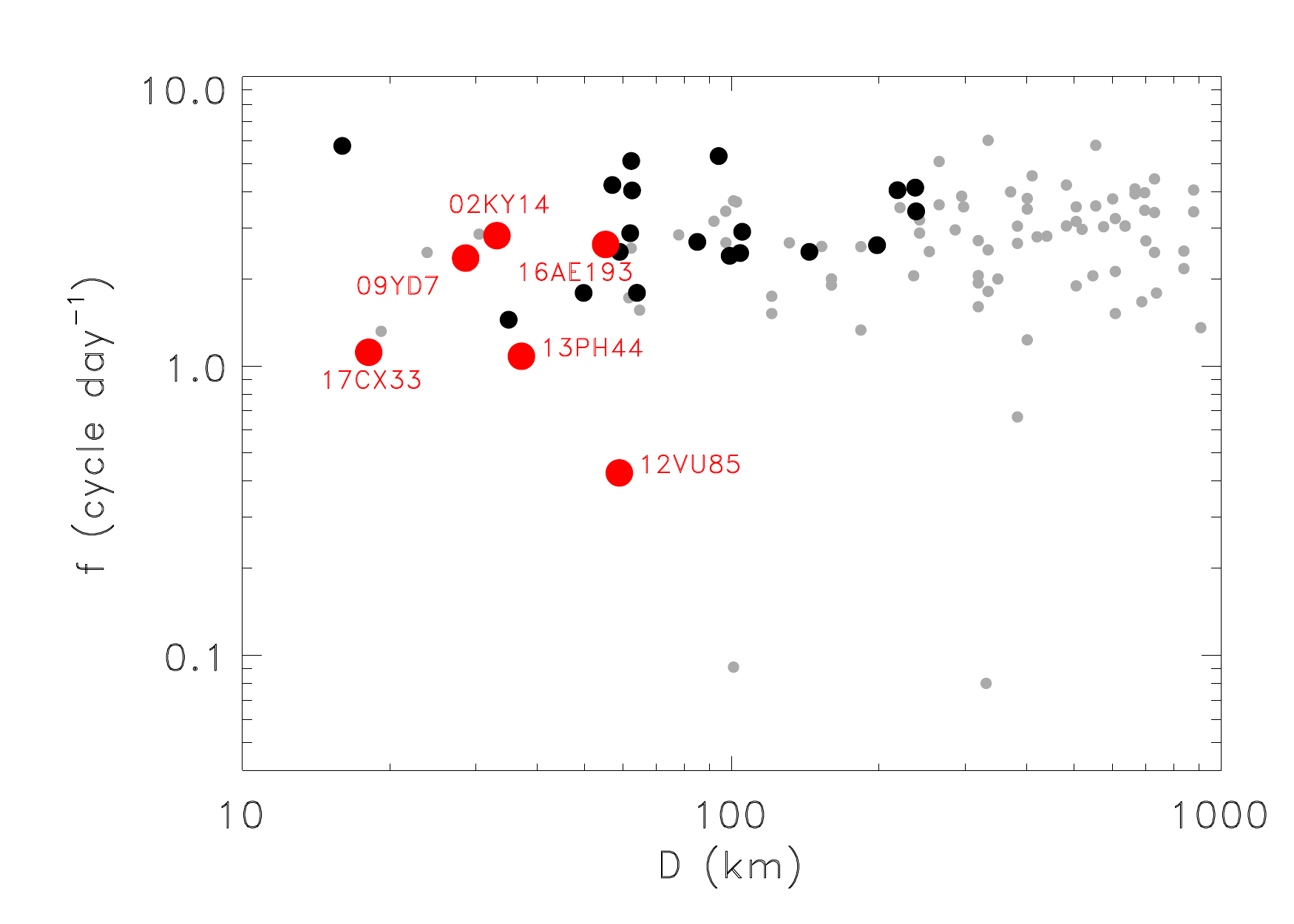}   
\includegraphics[width=0.75\textwidth]{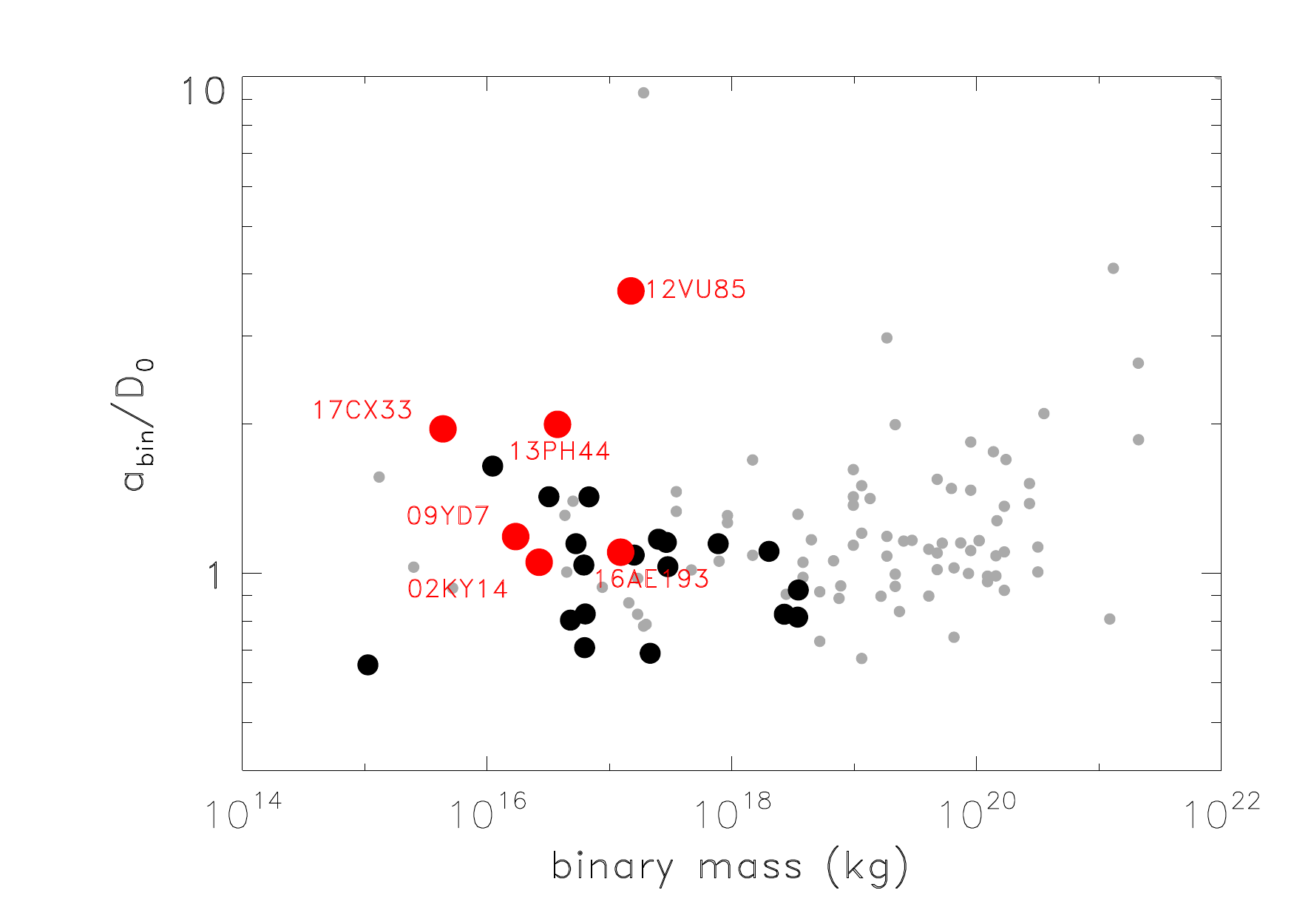} 
\caption[]{Upper panel: Rotational frequency (cycle\,day$^{-1}$) versus the estimated diameter; Lower panel: $a_{bin}/D_0$ binary semi-major axis to size ratio versus the binary mass estimated. On both panels black and gray dots correspond to Centaurs from the reference sample and TNOs, respectively. Targets investigated in this paper are marked by red symbols. }
\label{fig:binarity}
\end{figure}

We present the rotational frequency (cycle\,day$^{-1}$) as a function of the estimated size in Fig.~\ref{fig:binarity} (upper panel), and compare it with other Centaurs (black dots) and with the population of transneptunian objects (TNOs), the latter ones taken from the LCDB. As seen previously in the rotation period comparison, the rotational frequencies of our targets are typically lower than those of other Centaurs and TNOs.   

We used the ratio of $a_{bin}$ to the effective diameter $D_0$ of the two equally sized bodies to characterise the potential binarity { \citep[see][for a more complex tidal evolution analysis using this parameter]{Taylor2011}}. For having enough space for two bodies in such a system $a_{bin}/D_0$\,$>$\,1 has to be fulfilled ($a_{bin}/D_0$\,=\,1 corresponds to a contact binary). As shown in Fig.~\ref{fig:binarity} 1\,$<$\,$a_{bin}/D_0$\,$<$\,2 for many slower rotating Centaurs, but there are no objects with $a_{bin}/D_0$\,$\geq$\,2 in the Centaur reference sample. 
{ Note that 'classical' binary systems with tidal locking do not appear in these plots, as their rotational/orbital periods are notably longer (several days) than the typical rotation periods observed from light curves. These known binary systems also have notably larger separations than that can be deduced for a typical light curve target.}
The same calculations were performed for our targets. Radiometric size estimate is available for \ky\, only \citep{Duffard2014}, in the other cases we used our calculated R-band absolute brightness ($m_{110}^R$), and assumed a specific colour to obtain the H$_V$ V-band absolute magnitude. The colour distribution of Centaurs is bimodal \citep[e.g.][]{Peixinho2012,Peixinho2015} and the two colour groups correspond to two different average albedos \citep{Lacerda2014,Farkas2018}. For our targets we have colour information for \vu\, and \ky\, \citep{Tegler2016,Wong2019}, but as mentioned above, \ky has a reliable size estimate from radiometry. For \vu\,  \citet{Tegler2016}
obtained V-R\,=\,0.63$\pm$0.04\,mag, and with this colour \vu\, is in the 'bright-red' group identified by \citet{Lacerda2014} which has a mean albedo of \geomalb\,=\,0.16$\pm$0.08. We used this value to obtain the effective diameter of \vu\, from the absolute magnitude. As we have no colour information for the other four targets we used a mean V-R\,=\,0.558\,mag and p$_V$\,=\,0.088, obtained from the Centaur sample with known geometric albedos \citep{Duffard2014,Farkas2018}, averaged over the two colour groups. The lack of colour information introduces a V-R error of $\sim$0.18\,mag in the H$_V$ estimate \citep{Peixinho2015}. 

Interestingly, three of our targets, \ph\, \cxshort\, and \vu\, have $a_{bin}/D_0$\,$\geq$\,2, exceeding the values of the slowest rotating Centaurs, and also our other three targets have 1\,$\lesssim$\,$a_{bin}/D_0$\,$\lesssim$\,2, in the same range as the slower rotating Centaurs. Based on this estimate our six targets with rotation periods   {might be considered} as potential binaries { concerning this parameter only}. However, as discussed above, a distorted rotating body { or albedo variegations} may also be plausible explanations for \ky, \yd\, and \aeshort. 

\subsection{Tidal evolution timescales}

The simple calculations above assumed that the observed light curve period were both the rotation period and the orbital period of the binary, i.e. the system was tidally locked. It is, however, an important question whether the rotation of the individual bodies could have been slowed down by tidal forces and synchronized to the orbital period. Tidal dissipation is governed by the internal structure and composition of the bodies, and is usually considered through the tidal dissipation factor $Q$ \citep[e.g.][]{GS1966}. $Q$ factors of the terrestrial planets and satellites are usually found to be in the 10\,$\leq$\,$Q$\,$\leq$\,500 range, and for our calculations in the following we apply the generally accepted $Q$\,=\,100. However, as it is discussed e.g. in \citet{Grundy2007}, smaller objects require a correction to $Q$, since their rigidity can be large compared with their self gravity, leading to deformations smaller than expected in hydrodynamic equlibrium. Therefore we use a corrected tidal dissipation factor, $Q'$, obtained as
\citep[same as eq.~4 in ][]{Grundy2007}:
\begin{equation}
Q' = Q\bigg(1 + {{19\mu}\over{2g\rho R}}\bigg)
\end{equation}
where $\mu$ is the rigidity, $g$ the gravitational acceleration on the surface, $\rho$ the density and $R$ the radius of the object. This correction is very significant for small bodies with relatively rigid interiors. E.g. \citet{Grundy2007} obtained $Q'$\,=\,300--3$\cdot$10$^6$ using Q\,=\,100 for the Ceto-Phorcys system, in which the two bodies were in the 100-200\,km size range. For our small Centaurs this correction is even more significant. Assuming $\mu$\,=\,4$\cdot$10$^9$\,Pa rigidity (that of icy bodies, see e.g. Gladman et al., 1996) we obtain $Q'$\,$\approx$\,10$^7$ for all of our targets (actual values are listed in Table~\ref{table:tidal}). 

One of the important timescales related to the tidal evolution of binary systems is the orbit circularization timescale that we estimate as \citep{Noll2008}: 

\begin{equation}
\tau_{circ} = {{4Q'M_2}\over{63M_1}} \bigg( {{a^3}\over{G(M_1+M_2)}} \bigg)^{1/2} \bigg( {{a}\over{R_2}} \bigg)^5
\end{equation}

where $a$ is the semi-major axis of the binary orbit, $M$ is the mass, $R$ is the radius of the body and $G$ is the gravitational constant. The indices 1 and 2 refer to the primary and secondary, however, in our simple calculations all bodies are considered to be equal. 

For our binary systems $\tau_{circ}$ obtained through these calculations are in the order of 10$^4$--10$^6$\,yr, using the present estimated parameters of the systems, significantly smaller than the age of the Solar System. 
Another important question is whether in these systems the individual bodies could keep at least some of their own spin angular momentum and rotate with a period different from that of the binary orbit, or are fully spin locked due to tidal effects. We estimate this despinning (or spin-locking) timescale following \citep{Grundy2007}, as: 
\begin{equation}
\tau_{spin} = {{Q'\Delta \omega_1 M_1a^6}\over{G M_2^2 R_1^3}} 
\label{eq:despin}
\end{equation}

where $\Delta\omega_1$ is the change in angular speed with respect to the initial value. When the spin locking state is reached { the mean motion $n$ of the binary orbit is assumed to be equal to the angular speed obtained from the light curve period, i.e.  $\Delta\omega_1$\,$\approx$\,$\omega$\,=\,$n$. } For our targets the despinning timescales are 10$^5$--10$^7$\,yr, typically a few times longer than the corresponding circularization timescale. This suggests a fast tidal evolution for basically all systems, on timescales much smaller than the age of the Solar system (see also Table~\ref{table:tidal}).    


Kozai cycle tidal friction \citep{Porter2012} is a mechanism that can also create tight systems from an originally wider system with high eccentricity, if the inclination of the binary orbit to the heliocentric orbit is sufficiently large.  

\begin{table*}
\begin{tabular}{cccccccc|c}
\hline
             target  &  mass    &  R$_0$  &    g          &  $a$  & $Q'$  & $\tau_{circ}$ & $\tau_{spin}$ & $\rho_\mathrm{JE}$\\
                     & (kg)     &  (km)   & (cm\,s$^{-2}$)  & (km)  &       &  (yr)         &   (yr)    & (g\,cm$^{-3}$)\\
\hline
\ky & 2.7E+16 & 16.6   &  0.6  &  34.8  & 5.1E+07 &  2.0E+04 &  1.4E+05 & 0.540 \\ 
\yd & 1.7E+16 & 14.3   &  0.6  &  33.8  & 6.8E+07 &  5.9E+04 &  3.3E+05 & 0.388 \\
\vu & 1.5E+17 & 29.5   &  1.2  & 137.4  & 1.6E+07 &  1.1E+06 &  1.7E+06 & 0.018 \\
\ph & 3.8E+16 & 18.6   &  0.7  &  74.1  & 4.0E+07 &  1.0E+06 &  2.1E+06 & 0.088 \\
\aeshort & 1.2E+17 & 27.6   &  1.1  &  60.8  & 1.8E+07 &  9.9E+03 &  6.4E+04 & 0.480 \\
\cxshort & 4.4E+15 &  9.1   &  0.4  &  35.4  & 1.7E+08 &  3.8E+06 &  7.8E+06 & 0.094 \\
\hline           
            \gx &  1.6E+17  &   29.9   &    1.2  &  --  & 1.6E+07 &  -- & -- & -- \\
            \jj &  9.7E+17  &   54.8   &    2.1  &  --  & 4.6E+06 &  -- & -- & -- \\
       \plshort &  1.9E+17  &   31.7   &    1.2  &  --  & 1.4E+07 &  -- & -- & -- \\
       \fzshort &  8.0E+15  &   11.1   &    0.4  &  --  & 1.1E+08 &  -- & -- & -- \\

\hline           
\end{tabular}
\caption{Estimated binary system mass, binary size, surface gravitational acceleration, binary orbit semi-major axis, tidal dissipation parameter, and circularization and despinning timescale for our targets (see the text for the details of the estimation). We also list the ratio of the estimated semi-major axis, $a$, to the maximum semi major axis (a$_{max}$) for which despinning of the components can be expected within the lifetime of Solar system, 4.5$\cdot$10$^9$\,yr.  
The semi-major axis of the binary orbit and the tidal dissipation timescales cannot be estimated for targets without a known rotation period (bottom lines). Note that the system mass estimated for a binary is a factor of $\sqrt{2}$ {\it smaller} than it would be for a single object.
In the last column we list the estimated density assuming a single body with a shape of a Jacobi ellipsoid, considering the actual double peak light curve period and the observed light curve amplitude (also not obtained for targets without a known rotation period). 
\label{table:tidal}}
\end{table*}

\subsection{Encounters with giant planets}

Tenuously bound binaries may be disrupted by giant planet encounters and Centaurs are especially susceptible in this respect. For our assumed system configurations, however, the ratio of the calculated binary orbit semi-major axis to the Hill radius is $a_{bin}/r_H$\,$\leq$\,0.0025 for all our targets, while the Hill radii themselves are $r_H$\,$\leq$\,0.001\,AU. Encounters that close should be extremely rare \citep{Noll2006}. 
Concerning the target with the largest $a_{bin}$ in our sample, \citet{Wlodarczyk} investigated the dynamics of \vu, including close encounters with the giant planets Uranus and Neptune. According to their analysis this Centaur has no encounters with Uranus closer than $\sim$4\,AU, and with Neptune closer than $\sim$1\,AU. For 10\% of the closest encounter distance with Neptune (0.1\,AU) the gravitational influence distance of Neptune (Hill radius) would be $\sim$10$^4$\,km, significantly larger than the estimated $\sim$140\,km semi-major axis of the system, i.e. the binary system can be safely kept during these encounters (due to the similar mass and larger distances the encounters with Uranus are even safer). 

\section{Conclusions}

{ We have presented Kepler Space Telescope light curve measurements of ten Centaurs, observed in the course of the K2 mission. We were able to derive rotation periods for six of these targets, of which five are new period determinations. Three of our six targets fall in the P\,$\geq$\,20\,h regime,
not seen previously in ground based light curve period studies of Centaurs.}

{ Due to the low amplitudes the light curves of \ph\, and \ky\, can be explained either by albedo variegations, binarity or elongated shape. \yd\, and \aeshort\, are just above the amplitude limit and have relatively short rotation periods indicating that their light curves could be caused by elongated shape. Due to their slow rotations and higher light curve amplitudes \cxshort\, and \vu\, are the most promising binary candidates. }

Due to the lack of suitable spatial resolution by the current astronomical instrumentation binary systems in the typical distances of Centaurs cannot be discovered by direct imaging, but light curves with long rotation periods may be an indication for such systems. As shown for Centaurs in this paper and also previously for other small body populations \citep[e.g.][]{Szabo2017,Molnar2018} long, uninterrupted time series photometry is usually necessary to   {fully characterise} such systems. The K2 mission of the \textit{Kepler} was an excellent tool for these kind of studies \citep[see][for a summary]{Barentsen}. Similar results are expected from the TESS mission for Solar system targets \citep{Pal2018}.  


\section*{Acknowledgements}
The research leading to these results has received funding from the European Unions Horizon 2020 Research and Innovation Programme, under Grant Agreement No. 687378; from the K-125015, PD-116175, PD-128360, and GINOP-2.3.2-15-2016-00003 grants of the National Research, Development and Innovation Office (NKFIH, Hungary); and from the LP2012-31 and LP2018-7/2019 Lend\"ulet grants of the Hungarian Academy of Sciences.
L. M. was supported by the Premium Postdoctoral Research Program of the Hungarian Academy of Sciences. The research leading to these results have been supported by the \'UNKP-19-2 New National Excellence Program of the Ministry of Human Capacities, Hungary. 
Funding for the \textit{Kepler} and K2 missions are provided by the NASA Science Mission Directorate. The data presented in this paper were obtained from the Mikulski Archive for Space Telescopes (MAST). STScI is operated by the Association of Universities for Research in Astronomy, Inc., under NASA contract NAS5-26555. Support for MAST for non-HST data is provided by the NASA Office of Space Science via grant NNX09AF08G and by other grants and contracts.  The authors thank the hospitality of the Veszpr\'em Regional Centre of the Hungarian Academy of Sciences (MTA~VEAB), where part of this project was carried out. { We also indebted to S. Benecchi and an anonymous reviewer for their comments which have helped to improve the paper. }


\end{document}